# A collaborative and pattern-based training approach to knowledge acquisition and decision-making during the design of Software Architectures courses: A case study

Wilson Libardo Pantoja Yepez[1][0000-0002-7235-6036] and Luis Mariano Bibbó[2][0000-0003-4950-3605] and Julio Ariel Hurtado Alegría[1][0000-0002-2508-0962]

[1] Universidad del Cauca, Colombia
[2] Universidad Nacional de la Plata, Argentina
wpantoja@unicauca.edu.co

**Abstract.** This article describes a collaborative learning experience on Software Architecture (SA) between Universidad del Cauca (UNICAUCA) in Colombia and Universidad Nacional de la Plata (UNPL) in Argentina. The goal was to apply and evaluate training patterns, identifying effective practices for replication in other contexts. During the planning phase, both universities compared learning objectives, curricula, and teaching strategies to find common ground for improving student training. Selected training patterns were implemented, and their impact on professors and students was measured. As an integrating activity, a global development experience was carried out in the final part of the course, merging the work teams of the two educational institutions in a development iteration. The evaluation of this experience focused on the competencies achieved through the training patterns, their perceived usefulness, and ease of use based on the Technology Acceptance Model (TAM). The training addressed industry needs for software architecture design skills despite challenges such as the abstract nature of architectures, prerequisite knowledge, difficulty in recreating realistic project environments, team collaboration challenges, and resource limitations. A catalog of training patterns was proposed to provide quality training. These patterns help simulate industry-like environments and structure architectural knowledge for incremental learning. The ability to make architectural decisions is developed over time and through multiple project experiences, emphasizing the need for practical, well-structured training programs.
.

**Keywords:** Knowledge in Software Architecture, Training, Collaborative Teaching, Training Patterns, Knowledge Acquisition, Decision Making

## 1    Introduction

Software architects are responsible for proposing the structure of software applications to satisfy the quality attributes and constraints of the system and make the applications easily adaptable to business changes. According to Sherman and Unkelos-Shpigel [1], the role of the architect is responsible for design and technical decisions in the software development process and has the function of solving a problem by defining the structures of a system that can be implemented using certain technologies. However, finding the right balance between software attributes such as security, performance, usability, availability, maintainability, and interoperability, among others, takes a lot of work. These attributes can conflict, so the software architect needs to know tactics, patterns, and principles to help them make the right design decisions. Therefore, architects must define systems by applying abstract knowledge, proven methods, and technologies to create a solution that balances quality expectations. Thus, architects' duties also include soft skills, such as leadership, communication, and training [2]. For example, a software architect must understand stakeholder expectations and communicate decisions to designers, programmers, testers, installers, and other users of the system's architectural specifications.

For example, undergraduate engineering programs, such as Computer Science and Engineering, Information Technology, Software Engineering, and Systems Engineering, include many courses in which skills are developed, and technical knowledge related to building software is imparted through the use of programming languages and development platforms [3]. However, students in these programs need more knowledge of Software Architecture (SA) and design issues despite their growing importance to the software industry [4]. Recent graduates especially need more skills in proper design decision-making, practices, and knowledge related to software design in the business context[4].

The existing literature on SA education and practice points to different reasons for this lack of skills related to software architecture and design. First, there is a considerable gap between the academic perception of software architecture and its practice in the industrial sector [5]. Sometimes, the problems that industries consider most critical and challenging are not given due importance for research or university training [5]. Second, there is a gap between the skills that the industry expects from graduates in Software Engineering and the skills taught in the curricula [6]. Third, many institutions need a clear vision of the topics in which software architects should be trained [4].

Although numerous virtual platforms provide training in technological development environments and programming languages, no courses help to form the systemic and strategic thinking of a software architect [4]. This situation has meant that most people interested in SA issues must embark on a long self-study path.

Decisions around SA go hand in hand with information and communication technology issues. A software architect must have a broad knowledge of technologies to decide which platforms, frameworks, and languages favor certain technical and business qualities and constraints. The problem lies in the speed at which these technologies evolve. Along with the rapid development of information and communication technology (ICT), the job skills required by ICT industries are also changing rapidly. It becomes difficult for all stakeholders to assess the gap between their skills and those needed in such an evolving context. Although universities conduct periodic assessments of the curriculum, the time gap between the assessments and their updates makes it quickly outdated, as it cannot cope with the accelerated and rapid changes occurring in the industry and the environment [7]. Moreover, the problem is not only to know the gap but also to work on concrete strategies to reduce it.

Through a review, we sought to investigate in depth the reasons for the complexity associated with teaching SA. As a result, we found several causes of the difficulty involved in the training in SA topics at the undergraduate level [8]. We classified these causes by categories defined through the reading of the summaries: (1) Software industry, (2) Curricular contents, (3) Professors, (4) Nature of the architectures, (5) Resources, and (6) Students. Some of these challenges have many causes: the abstract nature of the archfitectures, the bases required by the trainees, the difficulty in recreating projects and environments in training contexts with similar characteristics to those of the industry, the challenges of working in teams, the lack of resources and updated contents, among others [9].

This paper describes a shared teaching experience by executing a case study that allowed us to evaluate the impact of applying some training patterns articulated in a guide called SAGITA (Software Architecture: GuIdeline for TrAining) in two SA courses: the National University of La Plata, Argentina, and the University of Cauca. This case study was selected because it was a representative case, i.e., two undergraduate Software Architecture courses in which the professors involved were willing to test SAGITA and the training patterns, motivated to improve the courses and align them with industry needs. The case study design was conducted under the guidelines Runeson and Host [10] proposed.

## 2      Related work

Giraldo et al. present and discuss the results obtained from a collaborative and distributed learning activity called CODILA (Collaborative and Distributed Learning Activity), which addresses the challenge of designing a software architecture for a communications infrastructure in a distributed manner [11].

Garousi et al. reviewed literature focusing on the challenges faced by software engineering graduates as they commence their professional journey, highlighting the mismatch between the skills acquired during their undergraduate studies and those demanded by the industry [12]. This examination enables educators and hiring managers to customize their educational and hiring strategies to equip the software engineering workforce better.

Niño and Anaya suggest a curriculum overhaul for the software engineering domain within systems engineering programs [13]. The primary outcome entails establishing a framework of professional competencies specific to software engineering, categorized into primary and secondary levels. Additionally, it delineates the foundational subjects crucial for fostering the competencies above.

Kiwelekar and Wankhede outline and categorize a series of learning objectives utilizing the Revised Bloom's Taxonomy (RBT) [14]. Their examination underscores the essential cognitive skills essential for architecture modeling. One of the critical advantages of categorizing learning objectives lies in the ability to align various educational processes, including instruction, learning, and assessment, more effectively, leveraging the framework proposed in this study.

Rupakheti and Chenoweth delineate a decade-long teaching experience within an undergraduate software architecture course integrated into a software engineering degree program [15]. They articulate their perspective on the practical objectives of instructing software architecture at this academic level. Emphasizing that the principal aim of the course is to equip students for advanced design scenarios encountered in the industry, they advocate for utilizing contemporary technologies and methodologies in student projects, along with rectifying architectural deficiencies in existing systems.

## 3      Training Patterns

This research presents a catalog of SA training patterns articulated in a guide called SAGITA (Software Architecture: Guideline for Training) that allows professors to design and execute courses at the undergraduate level. These patterns will enable the development of competencies in the creation, evaluation, and documentation of software architectures that align with the expectations of the software industry. The training patterns were extracted from the literature review, based on the experiences reported by professors proposing strategies to train their students with software architecture s.

The detailed description of SAGITA and the seven training patterns are available in PDF format on page SAGITA Final Version. In general terms, a "pattern" refers to a general, reusable solution to a recurring problem. It can be applied in various contexts, from architecture and design to problem-solving in specific fields, such as computer science, psychology, and biology. Christopher Alexander says: *"Each pattern describes a problem that occurs over and over again in our environment, and then describes the core of the solution to that problem, such that you can use this solution a million times, without ever doing it the same way twice"* [16].

Inspired by software patterns, design patterns, and software architecture patterns, we proposed the idea of training patterns [17]. Training patterns are reusable strategies for common problems in the design and execution of software architecture courses to achieve AS competencies. These patterns represent best practices and experiences by the teaching community and provide structured and efficient solutions for specific situ-ations. Each pattern describes a problem that occurs frequently in a given context and presents a general solution that can be adapted and applied to different scenarios. The seven training patterns we propose are:

1. **Mini-Projects-based training**. When students reach their first SA course, they have already seen programming and software development courses. At this level, students have skills in creating simple applications but working only with functional requirements. To develop a SA, the professional must consider quality attributes (scalability, performance, security, etc.) and satisfy functional requirements. Working with mini projects, students can practice architecture patterns and architecture tactics to favor the fulfillment of quality attributes.
2. **Large project-based training**. When students have acquired sufficient software development skills to attack medium to large development projects in complex domains, the professor could develop a course that connects the theoretical foundations of software architecture with practice through a large project of some complexity. Students learn SA through a complete real project where they can see the results of their architectural design as a product.
3. **Open-source projects-based training**. When recent graduates join the software industry, one of the initial challenges they face is developing software components on existing and usually large projects. Students work with an open-source project to have the unique opportunity to learn attitudes only present in real-world scenarios, which can increase their skills and confidence.
4. **In-house project-based training**. Despite the advantages of confronting students with modifying real systems, working with open-source projects can become too complex for students and professors. Therefore, an alternative is for professors to have their open-source application (In-house project) to teach SA with enough complexity of an industrial system.
5. **Cases-based training**. Making correct architectural decisions while constructing a software system is one of the significant skills a future software architect must develop. Case-based training focuses on designing and analyzing real software projects from companies. Students experience and use the theory and technology of SA design applied to specific projects to improve the teaching effect. Cases are usually taught through the flipped classroom.
6. **Problem-solving-based training**. Making architecture decisions as a team while constructing a new software system is one of the most critical skills a future software architect must develop. The most important decisions are made at the beginning of the creation of an application: technologies to be used, architecture patterns, and trade-offs between quality attributes, among others. After comes the development and maintenance of the application. It is an activity where the goal is to design the architecture of a system close to reality. The activity is usually done in teams, where each group is assigned an exercise to be solved in a given time.
7. **Games-based training**. Teaching SA is complex because the architect's role is multifaceted. The architect requires developing technical, analytical, and

> communication skills. Most talented architects have acquired extensive knowledge over many years of experience. Professors need fun teaching methods for shortened training time related to SA decision-making. Game-based training is an educational strategy that uses elements of games to foster student engagement, participation, and learning.

The competencies to be achieved by students in the SA field are very broad and complex. In this sense, we propose a list of the minimum competencies that the software industry expects from recent university graduates [18]:

- C01: Clearly identifies the relevant software quality attributes that will drive the architecture of a software system to be built.
- C02: Consistently design the software architecture by defining how components interact with each other.
- C05: Independently evaluates a software architecture to determine functional and non-functional requirements satisfaction.
- C08. Impartially performs a trade-off analysis to evaluate architectures.
- C11: Maintains existing systems and their architecture to achieve the evolution of software systems.
- C12: Redesigns existing architectures for migration to recent technologies and platforms.
- C18: Critically analyzes functional and quality attribute software requirements.
- C19: Understands business and customer needs quickly to ensure that requirements meet these needs.
- C22: Periodically performs reviews of the source code written by the development team.
- C23: Develops reusable software components.
- C28: Designs and implements test procedures considering architectural aspects (component/service types, integration).

## 4     Collaborative teaching strategy

Considering the challenges inherent in teaching SA to university students, professors from two academic institutions, Universidad Nacional de la Plata (Argentina) and Universidad del Cauca (Colombia), joined forces to establish collaborative activities, share resources, and employ joint pedagogical strategies. They initiated a course planning phase, during which we compared the learning objectives, curricula, and teaching techniques and strategies used by both universities to identify convergence points. As a result of this phase, a weekly planning of topics and activities was developed. During this planning, the following decisions were made for both courses:

- Work on a common project called Open Market (similar to the functionalities offered by Mercado Libre), in which the theoretical topics will be applied. The project had two important versions: the first one with a monolithic architecture and the second one a refactoring towards an architecture based on microservices.
- The project was worked on in teams of 4 and 5 students.
- Conferences with guests from the industry were planned to bring architectural design cases to real life.
- For the third part of the courses, a global development experience will be conducted between students from the two institutions.
- Recording and sharing classes between the two institutions.

## 5    Case study design, execution and evaluation

Below, we describe the steps we carried out during the study according to Runeson and Host (Runeson 2008).

### 5.1    Case study design

**Goal**. The objective of this case study was to obtain empirical evidence of the impact on the teacher and students of applying SAGITA and the training patterns to the two AS courses.

**Research questions**. The research questions we address are:
- RQ1: What impact did the training patterns have on students' competency development at the end of the courses?

- RQ2: In this case study, how did the professor evaluate the Guide's usefulness and ease of use?

- RQ3: How was the students' satisfaction with the course?

**Context of the case study.** This case study was carried out during the first academic period of 2023, and two institutions participated. The first was the Software Architecture Patterns elective course of the Faculty of Computer Science of the Universidad Nacional de la Plata, Argentina, with 14 students. The second is the Software Engineering II course of the Systems Engineering program of the Universidad del Cauca, with 26 students. The case study is an improvement since it seeks to improve certain aspects of the phenomenon studied, such as students' competencies in AS topics.

The two courses have common objectives:

- Learn to identify early on architecture decisions that, by definition, are costly to modify in the future.

- To understand the challenges and problems that give rise to the evolution of architectural.

- To know the advantages and disadvantages of each of the current architectural styles.

- Learn to contextualize each architectural style to choose the most appropriate one.

### 5.2    Preparation of data collection

The techniques used for data collection were surveys, document analysis, and semi-structured interviews[1]. Still, during the development of the conversation, the order can be decided. The interview questions applied to the professors at the end of the course were:

1. After reading the SA Course Design Guide, describe what changes you made to your course in terms of content, objectives, and methodology?

2. Was the guide useful to you in planning your AS course? Was it worth using? Did it save you time and effort, or, on the contrary, did it add more work?

3. Was the Guide's outline of the steps for designing an SA course easy for us to follow, or did it have complex steps to understand?

4. According to the SA course design Guide, which training patterns did you decide to apply in your course and why?

5. Explain how you implemented the training patterns (at what time of the semester, with what human and technical resources).

6. How did these patterns impact your students' skill development at creating, documenting, and evaluating SA?

7. What difficulties did you have applying the training patterns during the course development?

---

[1] In a semi-structured interview, the questions are planned, but they are not necessarily developed in the established order [10]

8. Do you consider that the training patterns you used to allow software architecture training to be closer to the real world of the software industry?

9. What needs to be added to the proposed Guide to fulfill its purpose of proposing a set of steps and strategies to support the design and implementation of software architecture courses at the undergraduate level?

### 5.3   Execution of the case study

The case study was carried out during the first academic period of 2023. Initially, course instructors were introduced to the SAGITA Guide. The guide was used to plan the course. In particular, it was used to define the topics to be developed, the resources to be used, the choice and application of the training patterns, and the definition of the class project. Finally, the topics to be taught in the two countries were organized, such as SOLID principles and architectural styles, emphasizing the microservices style.

In terms of human resources, due to the complexity of the course, a group of several professors was organized in each institution. In the case of Argentina, there was one teacher in charge of theory classes and two collaborators for development and deployment. In the case of the Universidad del Cauca, there was one teacher for the theory and another teacher for the laboratories.

Several training patterns were applied. At UNICAUCA, the *Mini-Projects-based training* training pattern was chosen as the central strategy. UNPL applied the *Large Project-based Training* pattern focused on migrating a monolithic application to a micro-services-based solution.

The two universities applied the *Cases-based training* pattern through the variant of lectures with invited software architects from the industry. Several lectures were organized throughout the courses, and each guest speaker presented the details of an actual architectural case: the context, the problem, the decisions made, and the results. In this way, the students learned from the experience lived by the architects. The three conferences were:

1. E-sidif project of the Ministry of Economy of Argentina.

2. Case of the chargeback module of the Mercado Libre application, migrating a monolith to microservices.

3. Case of large-scale architecture at Banco Galicia.

In the third part of the course, a *global development* experience was carried out as an integrating activity, merging the work teams of the two educational institutions in a development iteration. The planning made it possible to synchronize the courses in Colombia and Argentina so that they would arrive simultaneously at the third cut with the necessary knowledge and skills.

The two universities worked on the same class project, an E-commerce application similar to Mercado Libre called OpenMarket. However, UNICAUCA students developed a simpler version of the project to adapt it to mini-project development.

One helpful aspect was to record the classes and lectures on a web platform. This recording allowed students and professors in both countries to access the theoretical and practical content anytime.

At the end of the course, an interview was conducted with the professors of the two institutions.

### 5.4     Results analysis

We could answer the research questions posed by triangulating the data collected through interviews, surveys, and document analysis.

**RQ1: What impact did the training patterns have on students' competency development at the end of the courses?**

The Large Project-based Training pattern allowed students to put AS theoretical knowledge into practice by developing a large and complex project. The Open Market project was developed in teams of 4 and 5 students, who performed requirements engineering, proposed system architecture, made design decisions, and communicated effectively for implementation and testing. Although the software did not have real customers, the requirements were inspired by modern e-commerce applications such as Mercado Libre.

Overall, the implementation of this development project had a positive impact on the students. The course development revolved around this project. Relevant aspects of this case study are described below.

- *Practical experience*. Students gained hands-on experience working on this large project, which allowed them to apply the theoretical concepts learned in a practical environment and face real challenges encountered in large-scale software development..

- *Teamwork*. This software project required collaboration among team members. Students learned to work as a team, communicate effectively, and coordinate their efforts to achieve common goals. This skill is crucial in the actual software industry.

- *Communication and collaboration*. This project involved students taking on different roles as developers, designers, architects, testers, and database designers. Students had to communicate and collaborate with their professors and advisors, reflecting the working reality of the industry.

- *Complex problem-solving*. The software project presented complex technical and design challenges. Students addressed problems early in the project and made critical architectural decisions. This situation enhanced their problem-solving and decision-making skills.

- *Planning and management*. The project required careful planning and adequate time and resource management. Students gained skills in time estimation, task allocation, and resource management, which are fundamental to project management in the industry.

- *Understanding architecture*. By working on this significant project, students had the opportunity to experience first-hand how scalable and robust software architectures are designed and built. They saw how components are organized, dependencies are managed, and modularity is ensured.

- *Work with current technologies*. The project involved the use of various current technologies and tools, such as Java EE, Spring Boot, Postman, Docker, etc. Students became familiar with modern development tools and practices used in the industry.

- *Preparation for working with industry*. By facing challenges and situations similar to those they will encounter in their future jobs, students prepare to transition to the software industry after graduation.

- *Personal Portfolio*. These large, complex projects can become valuable additions to students' portfolios, helping them stand out when seeking employment after graduation.

- *Cultural diversity and perspectives*. The global development experience in the final part of the course between UNPL and UNICAUCA students had additional impacts. Working with students from different cultures and academic backgrounds provided students with the opportunity to gain a deeper understanding of cultural diversity and other ways of approaching problems. Students learned to collaborate in teams distributed geographically and across time zones. This situation enhanced their communication and collaboration skills globally, which is increasingly relevant in today's working world. However, the short iteration period made it difficult for students to collaborate with five teams; only one collaborated well and produced an integrated and functional software solution.

Thanks to applying the *Cases-based training* training pattern, it was possible to coordinate three talks with invited software architects from the industry. The students were able to see real cases from the industry. These talks had several positive impacts. Some of them were:

- *Industry linkage*. Industry guest conferences allowed students to connect directly with working professionals and gain a realistic understanding of how SA theoretical concepts are applied in the working world.

- *Practical application*. Learning about real-life software design cases allowed students to see how SA concepts become concrete and applicable solutions in real-world situations.

- *Variety of perspectives*. The guest speakers came from different companies and sectors, exposing students to various approaches. It broadened their perspective on the different ways in which design problems can be addressed.

- *Stories of success and failure*. Guests from the industry shared stories of successful projects and challenges and failures in software design. These stories provided valuable lessons on what works and what doesn't, which helped students make better decisions in their future roles.

- *Direct interaction*. Students could ask direct questions to industry professionals and get answers based on actual, hands-on experience.

Regarding the competencies achieved at the end of the course, each student was asked: *Evaluate to what degree this course helped you achieve each of the following competencies required by the software industry. The following scale will be used for all questions: 1. Not helpful, 2.* The answers can be seen in the Fig. 1 . We can see that some competencies such as C05 (*Independently evaluates a software architecture to determine the satisfaction of functional and non-functional requirements*), C08 (*Impartially performs trade-off analysis to evaluate architectures*) and C011 (*Easily maintains existing systems and their architecture to achieve the evolution of software systems*) students state that the course "helped a lot" to develop them in a large proportion. On the other hand, there are competencies that were not developed, such as C22 (*Reviews periodically the source code written by the development team*), since a large percentage of students state that the course "helped a little".

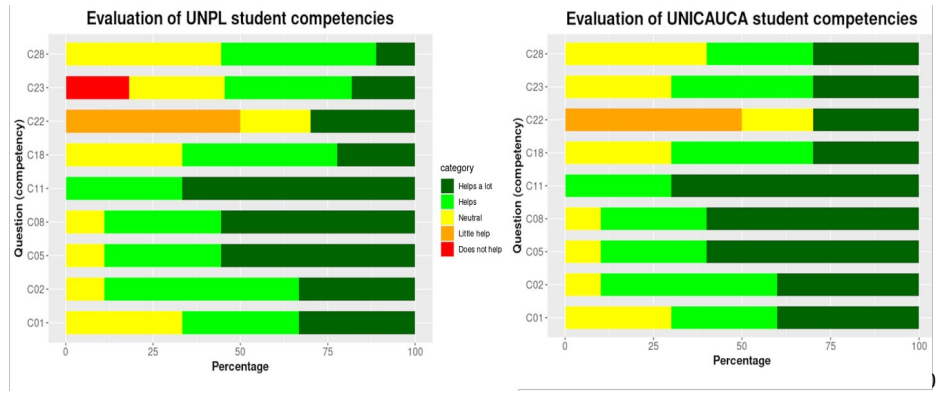

**Fig. 1.** Responses to the question: Please rate the extent to which you believe the Architectural Patterns course helped to achieve each of the following competencies? [Source: Own elaboration]

The main difficulties encountered in applying training patterns lie in the additional effort and time required by the professor. Each of these patterns entails a specific planning and execution phase. For example, coordinating talks with guests from the industry needed the professor to carry out tasks such as finding the guests, holding prior meetings, and coordinating their availability, among other aspects. Executing a large-scale project involved considerable efforts in drafting, revising, improving, and adjusting the project. In addition, following up and monitoring the students' projects represented a significant challenge. The professors need to be supported by more collaborating instructors who can assist them with issues related to Java technologies, application deployment, and student follow-up.

**RQ2: In this case study, how did the professor evaluate the Guide's usefulness and ease of use?**

Regarding the perceived usefulness, there were responses such as: "the *guide was handy because it allowed making improvements to the course. Some AS topics were adjusted, the necessary resources were chosen, and the training patterns were chosen to bring the course closer to the industry. The planning made was fulfilled during the execution. The guide provides accurate and relevant information on the design of Software Architecture courses. The concepts and guidelines presented directly apply to my area of study*".

Regarding the ease of use, professors expressed things like: "*The ease of use of the guide is remarkable and has been an overall positive experience. The guide features clear navigation and a well-organized structure. I can easily move between sections and find the information I need without difficulty. The language used in the guide is clear and accessible.*".

**RQ3: How was the students' satisfaction with the course?**

Valuable comments about the course were captured through surveys. Some of them are described below:

- "It was a great course and a pleasant experience, and I am grateful to my classmates, the professors, and those who came to lecture. It rarely happens that people from other fields come to share their experiences".

- "I liked the course methodology, which involved forming small groups with a project to develop. I also liked the implementation of the pipelines and that we had to use Docker; I feel that they were very useful for real developments.".

- "This is the first time a faculty subject has invited industry professionals to talk.".

- "It was interesting to learn how problems in the industry are being attacked from the point of view of architectural design. These talks allow us to learn about these experiences; otherwise, the only way is to face the problem in an actual situation."

- "I believe that talks are a valuable opportunity to exchange ideas, share knowledge, and strengthen the connection between people."

Several questions were asked through a student survey regarding overall satisfaction with the course. A Likert scale was used, with 1 Not at all satisfied and 5 Very satisfied. Regarding the question, "In general, how satisfied were you with what you learned in the course? The answers in a box plot can be seen in the Fig. 1.Error: no se encontró el origen de la referencia The box plots depict the distribution of student satisfaction. The horizontal line in the middle of the box represents the median satisfaction levels. In this case, the median is 4. We can say that most of the responses are in the scores range between 3 and 5, with a median of 4. It suggests that overall satisfaction tends to be high, as most responses are at the high end of the 1 to 5 scale.

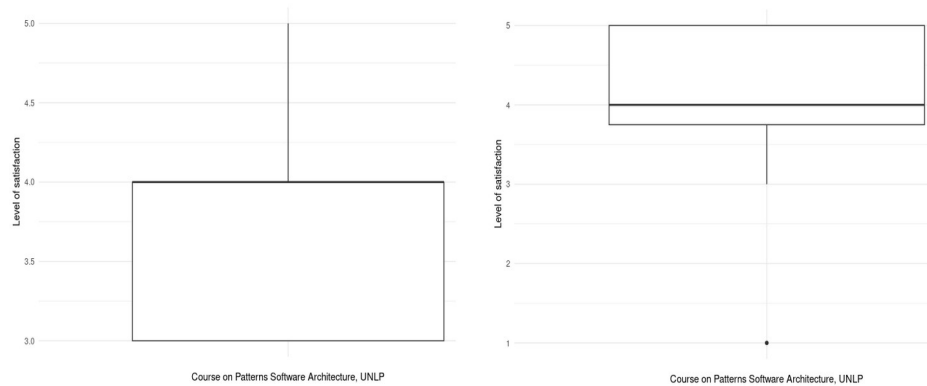

**Fig. 1.** Box plot of the question: How satisfied were you with what you learned in the course? [Source: Own elaboration]

## 6    Conclusions

An SA course aligned with industry needs is essential in the systems engineering curricula and related programs. However, training undergraduate students with the AS competencies demanded by industry has many challenges.

This research sought to organize knowledge that would help university professors design and conduct their AS courses in such a way that they would be able to face the difficulties of teaching AS and, on the other hand, align students' competencies with the needs of the industry. To this end, a catalog of seven training patterns was identified, structured, and documented, articulated within a guide that helped professors to design and execute courses at the undergraduate level that developed competencies in the creation, evaluation, and documentation of software architectures in line with the expectations of the software industry. These training patterns solve the problems that professors recurrently have to face, recreate the way of working in the classroom similarly to the environments used by the industry, helping the teacher to make decisions and, on the other hand, organize the architectural knowledge so that class activities can be structured and facilitate incremental learning. The ability to make architectural decisions is fundamental to the training process, but it takes time to develop and is accumulated through multiple project experiences.

Overall, the professor's satisfaction with using and applying SAGITA yielded positive results. In the case study, SAGITA helped professors plan their courses and implement strategies that brought the course closer to the reality of the software industry. In addition, we examined student satisfaction, evaluating through surveys the impact of the proposed strategies on their training. The results were generally satisfactory.

Using software *project-based training patterns* in teams provided UNPL and UNICAUCA students with a unique opportunity to develop practical skills, work in teams, face real challenges, and prepare for careers in the software industry. The pattern that involved bringing industry speakers into the classroom to share real-life software design cases enriched the students' educational experience by providing them with direct knowledge, practical insights, and valuable connections to the working world. The global software project development experience added a dimension of cultural diversity, international communication, and interdisciplinary collaboration, preparing students for an increasingly connected and globalized world of work. However, this activity for future courses should be improved in several aspects, both in time management and in carrying out previous activities to allow students to get to know each other. In addition, at the moment, there is no pattern of training in global software development.

Finally, the catalog of training patterns oriented professors on what and how to teach, developing the most relevant competencies for the current and future industry related to the creation, evaluation and documentation of ES in potential graduates of computer science programs.

**Acknowledgments.** A third level heading in 9-point font size at the end of the paper is used for general acknowledgments, for example: This study was funded by X (grant number Y).

**Disclosure of Interests.** It is now necessary to declare any competing interests or to specifically state that the authors have no competing interests. Please place the statement with a third level heading in 9-point font size beneath the (optional) acknowledgments, for example: The authors have no competing interests to declare that are relevant to the content of this article. Or: Author A has received research grants from Company W. Author B has received a speaker honorarium from Company X and owns stock in Company Y. Author C is a member of committee Z.

Contribution Title (shortened if too long)    17[5]     E. Lieh Ouh, B. Kok Siew Gan, and Y. Irawan, "Did our Course Design on Software Architecture meet our Student's Learning Expectations?," in *2020 IEEE Frontiers in Education Conference (FIE)*, Uppsala, Sweden: IEEE, 2020, pp. 1–9. doi: 10.1109/FIE44824.2020.9274014.

[6]     T. Akhriza, Y. ma, and J. Li, "Revealing the Gap Between Skills of Students and the Evolving Skills Required by the Industry of Information and Communication Technology," *International Journal of Software Engineering and Knowledge Engineering*, vol. 27, pp. 675–698, 2017, doi: 10.1142/S0218194017500255.

[7]     E. L. Ouh and Y. Irawan, "Applying Case-Based Learning for a Postgraduate Software Architecture Course," in *Proceedings of the 2019 ACM Conference on Innovation and Technology in Computer Science Education*, in ITiCSE '19. New York, NY, USA: Association for Computing Machinery, 2019, pp. 457–463. doi: 10.1145/3304221.3319737.

[8]     W. L. Pantoja and J. A. Hurtado, "Dificultades y retos de la formación de arquitectos de software en programas de pregrado: Causas y efectos," in *Tercer Simposio Doctoral del Doctorado en Ciencias de la Electrónica*, Popayán, Colombia., 2021. [Online]. Available: https://sites.google.com/view/simposiodoctoral2021/

[9]     M. Galster and S. Angelov, "What makes teaching software architecture difficult?," in *Proceedings - International Conference on Software Engineering*, Austin Texas: Association for Computing MachineryNew YorkNYUnited States, 2016, pp. 356–359. doi: 10.1145/2889160.2889187.

[10]    P. Runeson and M. Höst, "Guidelines for conducting and reporting case study research in software engineering," *Empir Softw Eng*, vol. 14, no. 2, pp. 131–164, 2008, doi: 10.1007/s10664-008-9102-8.

[11]    F. D. Giraldo *et al.*, "Applying a distributed CSCL activity for teaching software architecture," in *International Conference on Information Society (i-Society 2011)*, London, United Kingdom: IEEE, 2011, pp. 208–214. doi: 10.1109/i-Society18435.2011.5978540.

[12]    V. Garousi, G. Giray, E. Tüzün, C. Catal, and M. Felderer, "Closing the gap between software engineering education and industrial needs," *IEEE Softw*, vol. 37, pp. 68–77, 2020, doi: 10.1109/MS.2018.2880823.

[13]    M. Niño and R. Anaya, "Hacia un enfoque basado en competencias para la enseñanza de la ingeniería de software utilizando investigación-acción," in *Encuentro Internacional de Educación en Ingeniería ACOFI*, Cartagena de Indias, 21017. [Online]. Available: https://acofipapers.org/index.php/eiei/article/view/586

[14]    A. W. Kiwelekar and H. S. Wankhede, "Learning objectives for a course on software architecture," in *European Conference on Software Architecture*, 2015, pp. 169–180.

[15]    C. R. Rupakheti and S. V. Chenoweth, "Teaching Software Architecture to Undergraduate Students: An Experience Report," in *Proceedings - International Conference on Software Engineering*, Florence Italy: IEEE Press, 2015, pp. 445–454. doi: 10.1109/ICSE.2015.177.

[16]    C. Alexander, *A pattern language: towns, buildings, construction*. Oxford university press, 1977.

[17]    W. L. Pantoja, J. A. Hurtado, L. M. Bibbó, A. Fernández, and B. Ajay, "Towards a Software Architecture Training Pattern Language," in *Pattern Languages Of Programs*